\documentclass[aip,twocolumn]{revtex4-1}

\usepackage{graphicx}
\usepackage{tikz}
\usepackage{dcolumn}
\usepackage{bm}
\usepackage[utf8]{inputenc}
\usepackage[T1]{fontenc}
\usepackage{mathptmx}
\usepackage{etoolbox}
\usepackage{float}
\makeatletter
\def\@email#1#2{%
 \endgroup
 \patchcmd{\titleblock@produce}
  {\frontmatter@RRAPformat}
  {\frontmatter@RRAPformat{\produce@RRAP{*#1\href{mailto:#2}{#2}}}\frontmatter@RRAPformat}
  {}{}
}%
\makeatother
\usepackage{amsmath}
\usepackage{url}
\usepackage{xcolor,colortbl}
\usepackage{needspace}

\begin{document}
\onecolumngrid
\noindent\textcolor{gray}{This article may be downloaded for personal use only. Any other use requires prior permission of the author and AIP Publishing. This article appeared in J. Appl. Phys. 134 (9) and may be found at https://doi.org/10.1063/5.0157208.}

\title{High-sensitivity transition-edge-sensed bolometers: improved speed and characterization with AC and DC bias}

\newcommand{\caltech}{Department of Physics, California Institute of Technology, Pasadena, California 91125, USA}
\newcommand{\jpl}{Jet Propulsion Laboratory, California Institute of Technology, Pasadena, California 91125, USA}
\newcommand{\goddard}{Observational Cosmology Lab, Code 665, NASA Goddard Space Flight Center, Greenbelt, MD 20771}
\newcommand{\sron}{SRON, Netherlands Institute for Space Research, 2333CA Leiden, The Netherlands}

\author{Logan Foote}
\affiliation{\caltech}
\affiliation{\jpl}
\email[]{lfoote@caltech.edu}

\author{Michael D. Audley}
\affiliation{\sron}
\author{Charles (Matt) Bradford}
\affiliation{\caltech}
\affiliation{\jpl}
\author{Gert de Lange}
\affiliation{\sron}
\author{Pierre Echternach}
\affiliation{\jpl}

\author{Dale J. Fixsen}
\affiliation{\goddard}
\affiliation{CRESST/University of Maryland, College Park, MD 20742}
\author{Howard Hui}
\affiliation{\caltech}
\affiliation{\jpl}
\author{Matthew Kenyon}
\affiliation{\jpl}
\author{Hien Nguyen}
\affiliation{\caltech}
\affiliation{\jpl}
\author{Roger O'Brient}
\affiliation{\caltech}
\affiliation{\jpl}
\author{Elmer H. Sharp}
\affiliation{\goddard}
\affiliation{Alora Science \& Design LLC, 7007 Leesville Blvd, Springfield, VA 22151}
\author{Johannes G. Staguhn}
\affiliation{The William H. Miller III Department of Physics \& Astronomy, Johns Hopkins University, 3400 North Charles Street, Baltimore, MD 21218}
\affiliation{\goddard}
\author{Jan van der Kuur}
\affiliation{\sron}
\author{Jonas Zmuidzinas}
\affiliation{\caltech}
\affiliation{\jpl}

\date{\today}
\begin{abstract}
    We report on efforts to improve the speed of low-G far-infrared transition-edged-sensed bolometers. We use a fabrication process that does not require any dry etch steps to reduce heat capacity on the suspended device and measure a reduction in the detector time constant. However, we also measure an increase in the temperature-normalized thermal conductance (G), and a corresponding increase in the noise-equivalent power (NEP). We employ a new near-IR photon-noise technique using a near-IR laser to calibrate the frequency-domain multiplexed AC system and compare the results to a well-understood DC circuit. We measure an NEP white noise level of 0.8 aW/rtHz with a 1/f knee below 0.1 Hz and a time constant of 3.2 ms. \\ \\
\end{abstract}

\maketitle

\Needspace{40 in}
\clearpage
\section{Introduction}
\label{sec:introduction}
Transition-edge sensed (TES) bolometers and their SQUID-based readout systems have been the detectors of choice for many recent ground-based astrophysics and cosmology experiments in the trans-millimeter bands [ACT (2015)~\cite{thornton_atacama_2016,holland_mechanical_2016}, Bicep2~\cite{ade_bicep2_2014}, Bicep3~\cite{ade_bicepkeck_2022}, GISMO~\cite{staguhn_gismo_2014}, HAWK+~\cite{harper_hawc_2018}, Keck Experiment~\cite{kernasovskiy_optimization_2012},  PolarBear~\cite{collaboration_measurement_2020}, SPIDER~\cite{ade_constraint_2022}, SPT~\cite{sobrin_design_2022}, etc.].
\par TES bolometer arrays have generally demonstrated excellent systematic performance in deep microwave background measurements (BICEP-Keck/SPIDER Collaboration 2015~\cite{ade_antenna-coupled_2015}, SPT 2018~\cite{everett_design_2018}, BICEP-Keck Collaboration 2022~\cite{ade_bicepkeck_2022}, ACT~\cite{holland_mechanical_2016}).  They are, thus, a strong candidate for future space missions, which would have lower backgrounds, both in the CMB bands (EPIC\cite{bock_study_2009}, Litebird\cite{litebird_collaboration_probing_2023}) and also in the far-infrared (far-IR), where no commercial high-performance detector systems exist. The far-IR is particularly compelling for a space mission because of the very low natural backgrounds across a full decade in wavelength from 25 to 300 $\mu$m.  A cryogenically-cooled telescope in this band offers large potential scientific gains for a range of astrophysical applications from planet formation to galaxy evolution \cite{meixner_19,origins_report,Roelfsema_18,Glenn_21}, particularly via moderate-resolution ($\lambda / \delta\lambda \approx 100-500$) spectroscopy \cite{Bradford_21}.  However, fully capitalizing on the low-background platform requires sensitivity  matching the detector sensitivity to the photon shot noise in the narrow band - this results in a target noise equivalent power (NEP) of $0.1\;\text{aW}/\text{Hz}^{1/2}$ or lower.
\par The ultralow-noise requirement poses new challenges.  The devices must have low thermal conductance (G) in the suspension, while maintaining sufficient speed; therefore, fabrication techniques must maintain low heat capacity in the released devices.  Additionally, characterization is challenging at the very low loadings (of order 1 aW or less), since bolometers are susceptible to all forms of radiation; both the electronics and the thermal radiation must be heavily filtered.  Early pioneering work in low-G far-IR TES sensors was made at JPL \cite{kenyon_progress_2006,Beyer12,Beyer14,Kenyon_14}, where NEPs approaching $0.1\;\text{aW}/\text{Hz}^{1/2}$ were demonstrated with DC voltage bias \cite{Irwin_95,Irwin_97}, albeit with the speed of response of several milliseconds.  Subsequent work at SRON reported comparable sensitivity with speeds below 1 ms \cite{Audley_16,Khosropanah_16,Ridder_16,Suzuki_16}, using the SRON AC bias system \cite{van_der_kuur_small-signal_2011,HIjmering_16}.
\par Here, we report the development of a low-heat-capacity wet-release process for fabricating low-G far-IR TES sensors and characterization of these devices. We employ both AC and DC biasing schemes and present a novel absolute power calibration technique using near-IR (1550~nm) shot noise, which overcomes uncertainties in the electrical transfer through the system.  We find a two to four times improvement in device speeds relative to our previous dry-release devices, but thermal conductances and associated noise-equivalent powers are five to ten times higher than the previously published ones for devices with similar geometries \cite{Kenyon_14}.
\par This paper is organized as follows: following the introduction in Sec.~\ref{sec:introduction} is the array design and fabrication in Sec.~\ref{sec:design_fab}. Sec.~\ref{sec:experimental_setup} describes the experimental setup, which covers both the DC and AC setup, and the setup for the measurement of the noise using the near-IR laser. Sections~\ref{sec:ac_measurements} and ~\ref{sec:dc_measurements} describe the detail of AC and DC measurements, respectively. These measurements include time constants, optical responsivity, and noise. Sec.~\ref{sec:summary} is the summary.

\section{Array Design and Fabrication}
    \label{sec:design_fab}
    \begin{figure*}[t]
        \centering
        \includegraphics[width=6.2in]{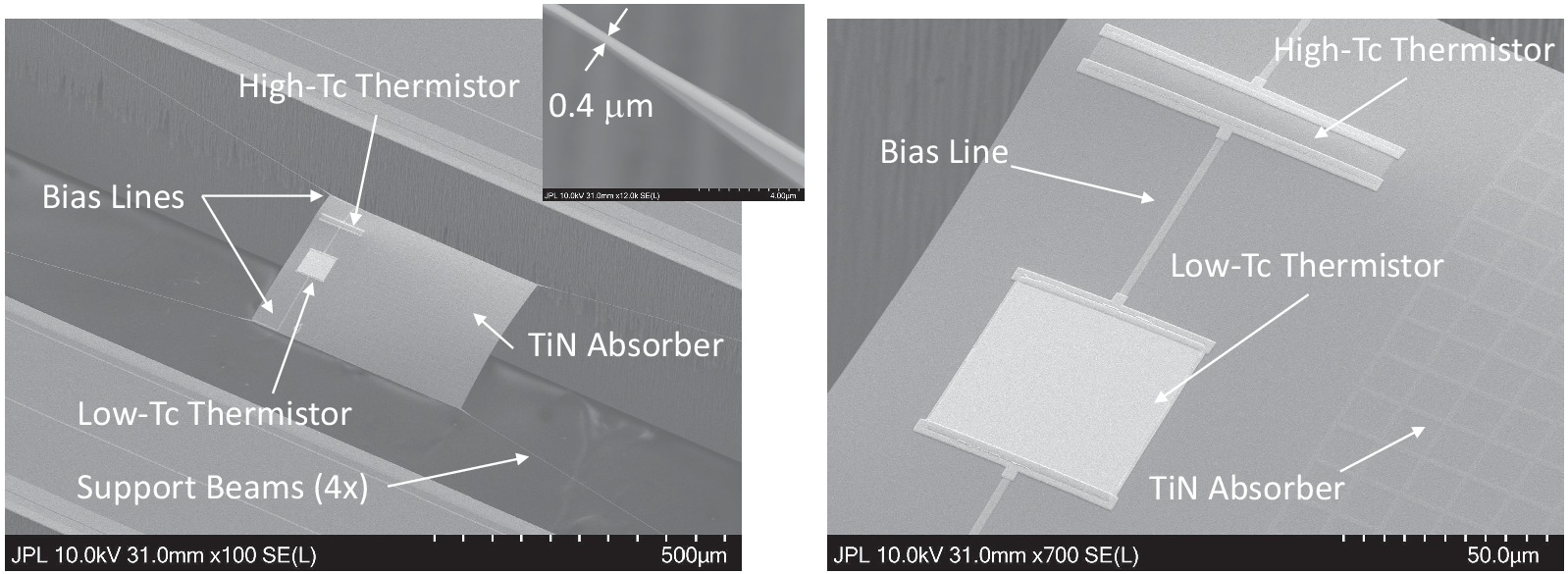}
        \caption{(Left) Micrograph of a single pixel.  Each pixel is a suspended low-stress silicon nitride (LSN) structure consisting of an absorber and four LSN support beams (see the inset) that connect the absorber to the substrate. (Right)  A dual-$\text{T}_c$ thermistor consisting of a 50 nm Ti/150 nm Au and 50 nm Ti is patterned on the absorber and electrically connected to niobium bias lines running along the support beams. A TiN absorber is patterned next to the thermistors.}
        \label{fig:tes_image}
    \end{figure*}

    \par Each TES sensor consists of a free-standing low-stress silicon nitride (LSN) membrane (0.25 x 400 x 300 $\mu\text{m}^3$) which connects to the substrate through four 0.25 $\mu\text{m}$ thick s 1000 $\mu\text{m}$ long s 0.4 $\mu\text{m}$ wide LSN support beams.  Niobium wires (30 nm thick) run along two of the support beams and electrically connect to an 80 x 8 $\mu\text{m}^2$ high-$\text{T}_c$ titanium thermistor (100 nm) in series with a 50 x 50 $\mu\text{m}^2$ low-$\text{T}_c$ bilayer consisting of titanium (50 nm) and gold (150 nm).  A grid of titanium nitride wires (1.5 $\mu\text{m}$ wide x 12 $\mu\text{m}$ long x 30 nm thick) with a $\text{T}_c$ of $\sim1\;\text{K}$ define the optical absorber, which has an effective sheet resistance of $\sim400\;\Omega/\text{sq}$.  Each TES sensor is arranged in an array with a format size of 5 x 12 with a column and row spacing of 3.73 and 0.99 mm, respectively. Micrographs of a single pixel are presented in Fig.~\ref{fig:tes_image}.
    \begin{figure*}[t]
        \centering
        \includegraphics[width=6.2in]{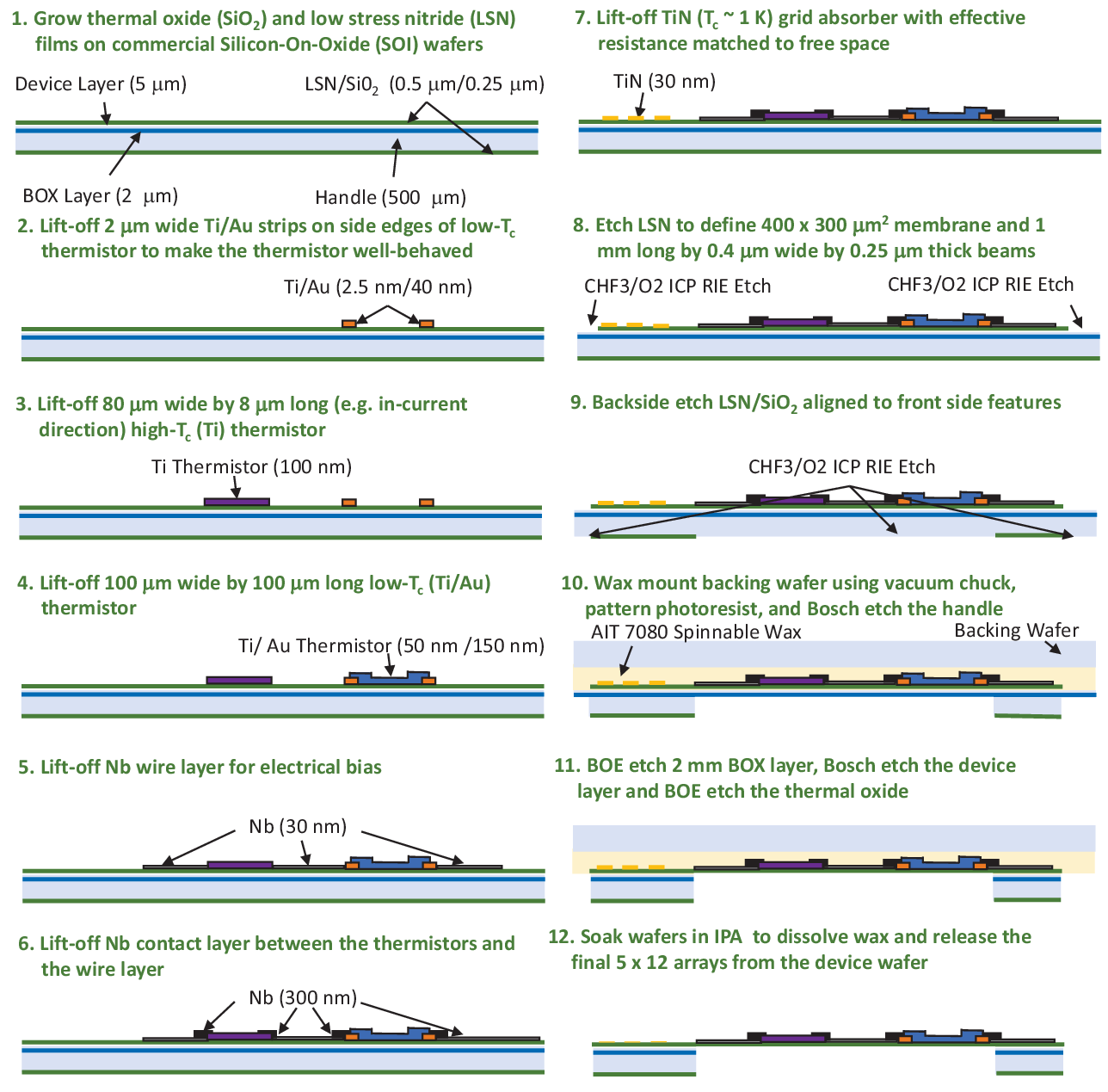}
        \caption{Diagram of the fabrication processes.}
        \label{fig:fabrication_flow}
    \end{figure*}
    \par Based on prior measurements, it was determined that dry etching the LSN membrane causes surface roughness on the membrane, which manifests as excess heat capacity \cite{beyer_characterizing_2010}. This excess heat capacity slows the response time of the TES sensor beyond the requirement.  Therefore, a fabrication process was developed that does not require any dry etch steps that etch the top and bottom surface of the LSN.  Key details of this process are presented in Fig.~\ref{fig:fabrication_flow} and described as follows: (1) 500 nm thermal oxide and 250 nm low pressure, chemical vapor deposition silicon nitride (LPCVD) (-200 MPa compressive stress) were grown on bare 100 mm silicon-on-insulator (SOI) wafers (500 $\mu\text{m}$ handle; 2 $\mu\text{m}$ buffered oxide layer (BOX); 5 $\mu\text{m}$ device) at the Jet Propulsion Laboratory's (JPL) Microdevices Laboratory. (2) Ti (2.5 nm)/Au (40 nm) strips (2 $\mu\text{m}$ wide) were patterned to passivate the edges of the low-$\text{T}_c$ thermistor so that its overall properties are well-behaved. (3) a thermistor consisting of Ti (100 nm) was patterned via dc sputtering and a lift-off technique. (4) A thermistor consisting of a bilayer of Ti (50 nm)/Au (150 nm) was patterned via dc sputtering and a lift-off technique.  (5) Nb wires (30 nm) to bias the two thermistors were defined using dc sputtering and a lift-off technique. (6) A Nb contact layer (300 nm) was sputtered on the edges of the thermistors to ensure good electrical contact between the bias lines and the thermistors. (7) A TiN absorber (30 nm) was patterned via dc sputtering with a $\text{T}_c\sim1\;\text{K}$. (8) The LSN membrane was patterned using a dry CHF3/O2 ICP RIE process into a 400 x 300 $\mu\text{m}^2$ membrane and four 1 mm long x 0.4 $\mu\text{m}$ wide x 0.25 $\mu\text{m}$ thick support beams. (9) The backside LSN/SiO$_2$ was etched using a dry CHF$_3$/O$_3$ ICP RIE process aligned to the front side membrane and support beams. (10) The device wafer was wax mounted to a 150 mm backing wafer using AIT 7080 spinnable wax and a vacuum mounting technique.  The handle of the SOI wafer was removed using a Bosch process and a photoresist to define openings underneath the suspended membrane and support beams. (11) The BOX layer was removed using a buffered buffered oxided etch (BOE).  The device layer was removed using a Bosch process.  The thermal oxide underneath the LSN membrane was removed using buffered BOE;  (12) The arrays were removed by soaking the device wafer and a 150 mm backing wafer in IPA using a fixture, which allowed the 12 x 5 arrays to separate from the backing wafer.

\section{Experimental Setup}
    \label{sec:experimental_setup}
    The TESs were tested in a BlueFors LD250 Dilution Refrigerator (DR) system \cite{ld_2019} at JPL. The temperature is controlled using a PID loop with a Lake Shore Cryotronics Model 372 AC Resistance Bridge \cite{lakeshore}. The thermometer and heater are mounted on the mixing chamber plate near the TES sample box. The temperatures presented in this paper are measured using the same thermometer as the PID loop. The TESs are biased with a Tektronix AFG 31000 Series Arbitrary Function Generator \cite{afg31000_2022}. The bias voltage is reduced with a voltage divider before entering the cryostat. The bias and readout lines in the cryostat are phosphor-bronze shielded twisted pairs. The cabling connects to the TES housing box shown in Fig.~\ref{fig:cryobox} using a 37-pin micro-D connector. The TES housing box and the mixing chamber plate are gold-plated copper, and the fasteners are brass to ensure good thermal contact while cold. The bias lines pass through a 20 MHz RC low-pass filter on a printed circuit board and are wirebonded into a chip containing a Magnicon 2-stage SQUID \cite{integrated_2022}. The SQUID is fed back and read out with Magnicon XXF-1 electronics \cite{xxf-1_2011}. The output of the SQUID electronics is recorded with a Pico Technology PicoScope 6403E \cite{picoscope_2022} USB oscilloscope. The bias and readout system is set up in either a DC or an AC mode depending on the measurement. The AC setup is described in Sec.~\ref{subsec:ac_setup} and the DC setup is described in Sec.~\ref{subsec:dc_setup}.

    \subsection{AC Setup}
        \label{subsec:ac_setup}
        \begin{figure}[h]
            \centering
            \includegraphics[width=3.1in]{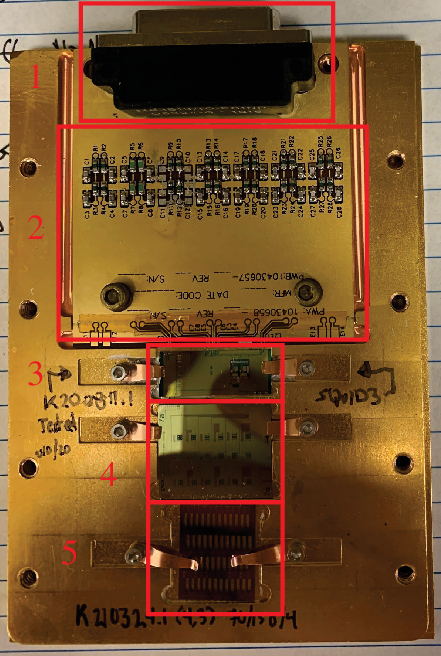}
            \caption{Photo of the TES housing box with the lid removed. The components are as follows: 1. 37-pin Micro-D connector, 2. RC filter circuit board, 3. Magnicon SQUID chip, 4. 18-channel LC filter chip, and 5. 60-pixel TES subarray (5 columns x 12 rows).}
            \label{fig:cryobox}
        \end{figure}
        \begin{figure*}[t]
            \centering
            \includegraphics[width=6.2in]{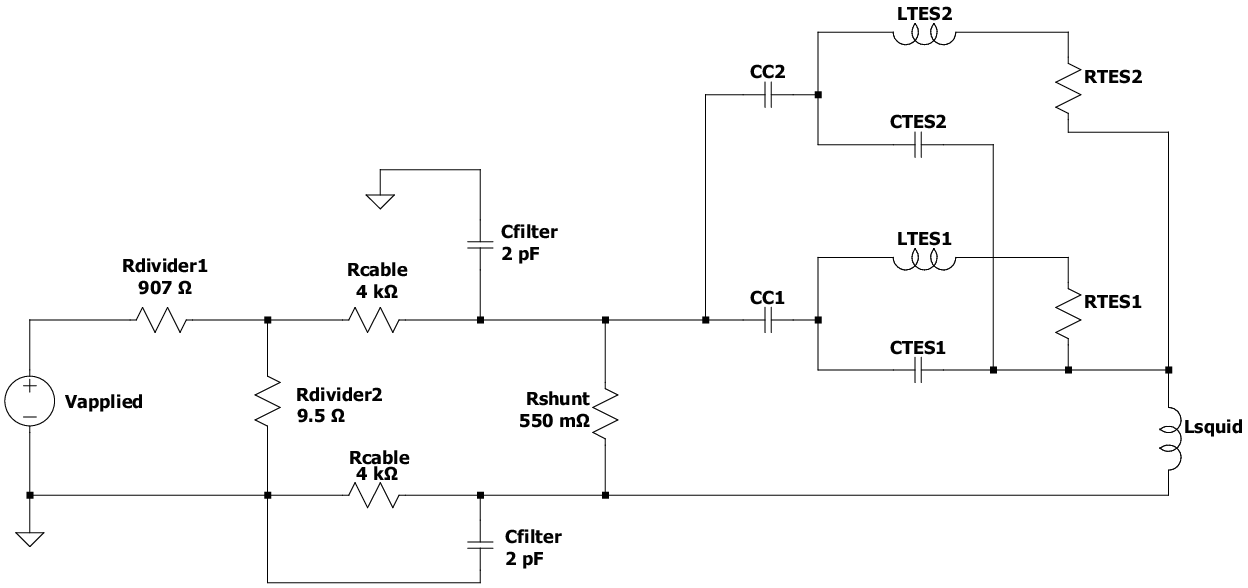}
            \caption{Spice model of the AC setup circuit. The components are as follows: the function generator (Vapplied), voltage divider resistors (Rdivider1 and Rdivider2), cables with added resistors (Rcable), filter capacitors (Cfilter), a shunt resistor (Rshunt), an LC chip (CC, CTES, and LTES), TESs (RTES), and the SQUID (Lsquid). Only 2 of the 18 LC channels are included in the circuit diagram.}
            \label{fig:ac_circuit}
        \end{figure*}

        The AC setup uses a cryogenic LC resonant filter array designed and fabricated at the Netherlands Institute for Space Research (SRON), developed as part of their AC multiplexing scheme\cite{vanderKuur04}, which has successfully supported 160 far-IR TES pixels in a single circuit \cite{holland_160_2014}.   Our work used an 18-channel test chip with resonance frequencies between 1 and 5 MHz (component 4 in Fig.~\ref{fig:cryobox}).   A circuit model of the AC system is presented in Fig.~\ref{fig:ac_circuit}. Each TES is wirebonded into one of the LC resonant circuits.  The 18-channels are connected to two TESs without absorbers, two TESs with absorbers, four thermistors, and 10 shorts.
        \par The MHz band of the AC bias circuit is beyond the closed-loop bandwidth of the SQUID and commercial electronics; therefore, the SQUID is operated in an open-loop mode for the AC setup.
        \par The SQUID is operated in the regime where the current to voltage conversion is linear to within 5\%.  The current scale is calibrated by measuring the voltage response of a single flux quantum, and using the reported input coil coupling for the SQUID from the Magnicon datasheet (measured at 320 mK).  Measurements that use the AC setup are presented in Sec.~\ref{sec:ac_measurements}.

    \begin{figure*}[t]
        \centering
        \includegraphics[width=4.2in]{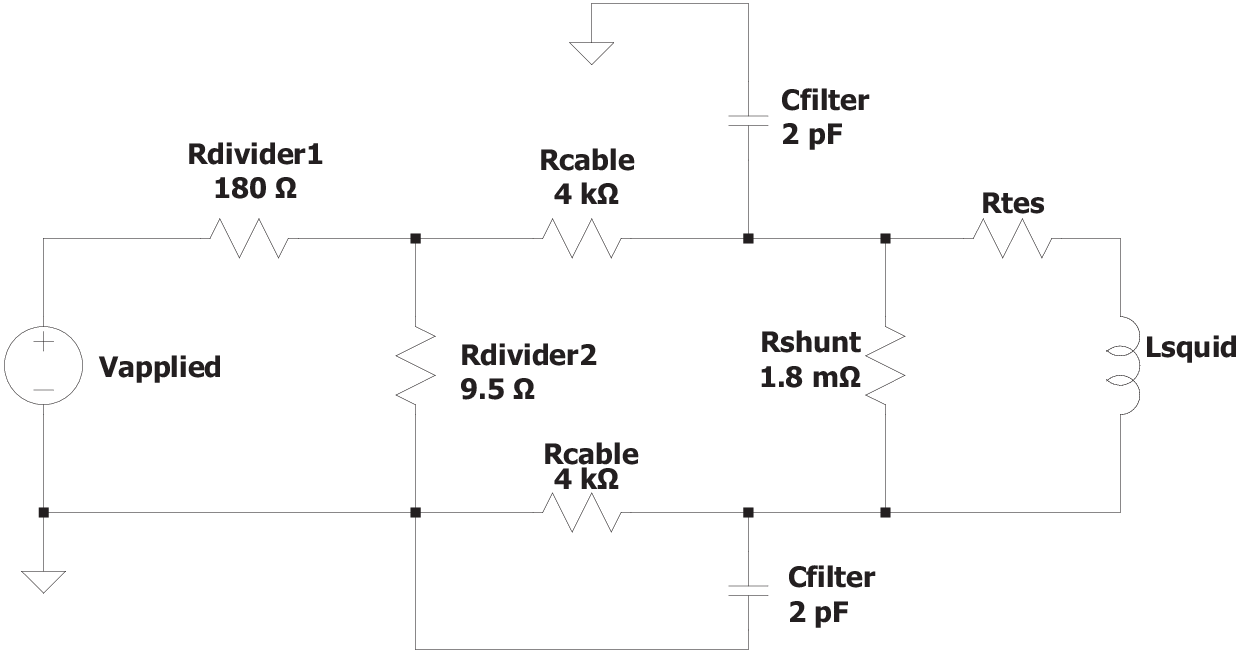}
        \caption{Spice model of the DC setup circuit. The components are as follows: the function generator (Vapplied), voltage divider resistors (Rdivider1 and Rdivider2), cables with added resistors (Rcable), filter capacitors (Cfilter), a shunt resistor (Rshunt), the TES (RTES), and the SQUID (Lsquid).}
        \label{fig:dc_circuit}
    \end{figure*}

    \subsection{DC Setup}
        \label{subsec:dc_setup}
        We complement our AC measurements with DC measurements, in which the LC chip is bypassed to measure a single TES with a more traditional DC bias.  The DC measurements provided an independent check on the absolute power scale for the device without the uncertainties in the AC measurement,  validating our photon-noise calibration approach presented in Sec.~\ref{subsec:photon_noise_derivation}
        The circuit model for the DC system is shown in Fig.~\ref{fig:dc_circuit}. The readout and function generator are the same in the DC and AC setup. The SQUID is operated with feedback for the DC measurements.  Measurements with the DC setup are presented in Sec.~\ref{sec:dc_measurements}.

    \subsection{Near-IR Laser Setup}
        \label{subsec:laser_setup}
\begin{figure}[h]
        \centering
        \includegraphics[width=3.1in]{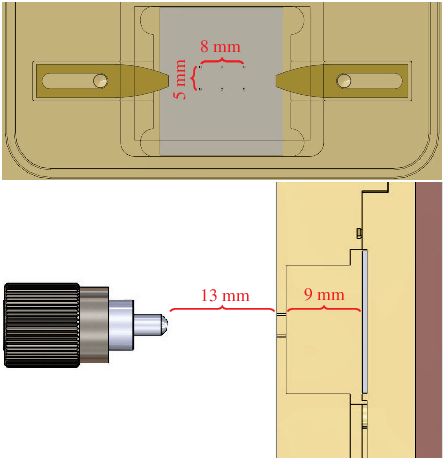}
        \caption{A SolidWorks model of the laser setup. Top: The TES housing box is shown with a transparent lid. The detectors are distributed along the detector wafer (gray). Bottom: Side view of the laser setup.}
        \label{fig:laser_setup}
    \end{figure}

        To overcome the uncertainties in the electrical power calibration of low-G TESs, we have developed a near-IR photon-noise calibration technique. For this measurement, six 0.5-mm diameter holes are drilled in the cover of the TES housing box over the detector array. The holes are drilled in an array of three (with 4 mm spacing) by two (with 5 mm spacing). The hole array is located above the four detectors that were tested with the AC setup. A fiber-optic cable with standard fiber connectors (FC connectors) is positioned about 1.3 cm in front of the holes, and the opening of the holes is 9 mm above the detector wafer. The FC connector is allowed to spray photons freely over the holes in the TES housing box. A CAD model of the laser setup is shown in Fig.~\ref{fig:laser_setup}. A ThorLabs S5FC1550S-A2 fiber-coupled super-luminescent diode (SLD) source\cite{thorlabs_sld} is used for the measurements presented in this paper.

\section{AC Measurements}
    \label{sec:ac_measurements}
    \par Two TESs with absorbers and two TESs without absorbers were studied sequentially with the AC system, all at a bath temperature of 100 mK. The resonance frequencies of the four LC filter channels associated with the TESs are given in Table~\ref{tab:tc_table}. Current vs voltage curves are measured for each TES, and bias voltages for the subsequent response and noise measurements are chosen such that the TES is at 1/2 the normal resistance. DC current vs voltage curves are presented in Sec.~\ref{subsec:dc_iv}.

    \subsection{Photon-noise Calibration}
        \label{subsec:photon_noise_derivation}
        \begin{figure*}[t]
            \centering
            \includegraphics[width=6.2in]{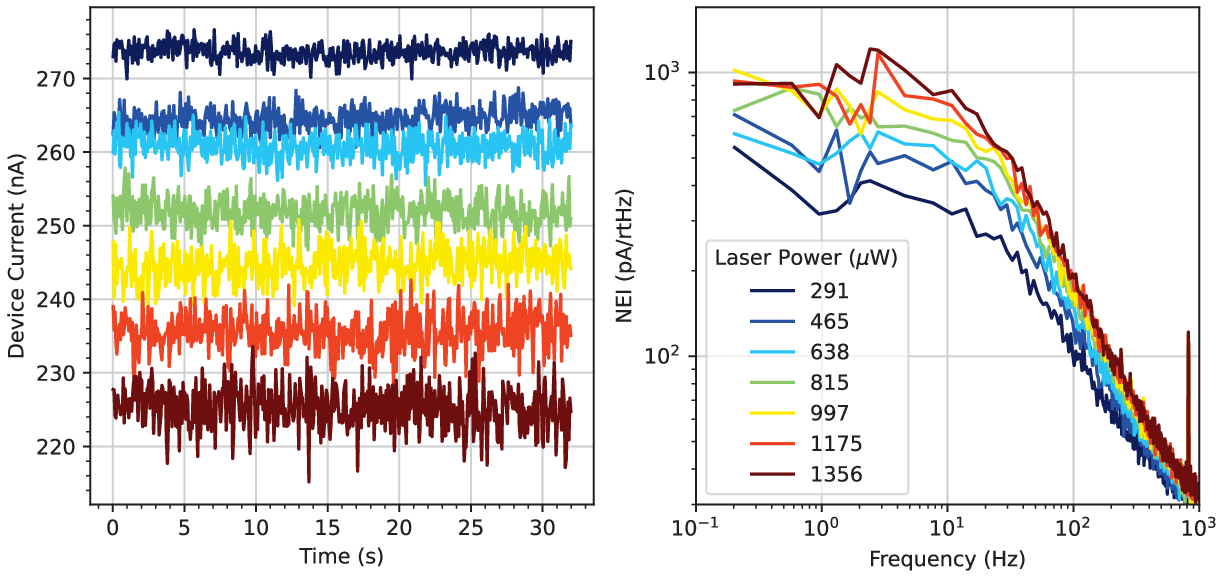}
            \caption{TES response to a laser power sweep. The laser power values are the measured laser power at the entrance to the cryostat. Left: TES current vs time. Right: NEI vs frequency. Both current and NEI are averaged in bins for visibility.}
            \label{fig:NEIvsI_raw}
        \end{figure*}
    \subsubsection{Approach}The absolute current and voltage scales in the AC setup are not known accurately due to uncertainty in the shunt resistor value, difficulty in establishing the open-loop MHz SQUID response, and reactive effects in the TES resonant circuit.
        To overcome these obstacles and provide absolute electrical power calibration, we have developed a technique using shot noise from a 1550 nm laser. When the laser is turned off, the current is $I_0$ and the NEI is $\text{NEI}_0$. Shot noise in the near-IR photon stream incident on the TES has a noise-equivalent power (NEP) of \cite{zmuidzinas_thermal_2003}
        \begin{align}
            \label{eq:photon_NEP}
            \text{NEP}_{\text{photon}} = \sqrt{2 P_{\text{photon}}h\nu_{\text{photon}}}.
        \end{align}
        Here $P$ is the photon power absorbed by the detector, $h\nu_{\text{photon}}$ is the photon energy, and the expression has units of $\text{W}/\text{Hz}^{1/2}$. The corresponding noise-equivalent current (NEI) generated by the fluctuating photon power absorbed by the TES is
        \begin{align}
            \label{eq:photon_NEI}
            \text{NEI}_{\text{photon}} = - S(\nu) \text{NEP}_{\text{photon}},
        \end{align}
        where $S(\nu)=dI/dP$ is the current to power responsivity. The frequency dependence of the responsivity is given by
        \begin{align}
            \label{eq:S_freq_dependence}
            S(\nu) = \frac{S(\nu = 0)}{\sqrt{1 + \left(2\pi\nu\tau\right)^2}},
        \end{align}
        where the rolloff is produced by the time constant of the TES, $\tau$.  The mean value of the current also changes as the near-IR power is introduced, creating an additional constraint
        \begin{align}
            \label{eq:current_shift}
            I - I_0 = S(\nu) P_{\text{photon}},
        \end{align}
       which is an excellent approximation for a voltage-biased TES so long as the responsivity is constant. The power of absorbed photons is eliminated by combining Eqs.~\ref{eq:photon_NEI} and~\ref{eq:current_shift} with Eq.~\ref{eq:photon_NEP}
        \begin{align}
            \label{eq:step_NEI}
            \text{NEI}^2_{\text{photon}} = 2 h\nu_{\text{photon}}S(\nu)(I-I_0)
        \end{align}
        where $\text{NEI}^2_{\text{photon}}$ is the photon shot-noise contribution to the NEI. The measured NEI is the quadrature addition of the photon NEI and the NEI with the laser off, $\text{NEI}_0$
        \begin{align}
            \label{eq:meas_NEI}
            \text{NEI}^2_{\text{measured}} = \text{NEI}^2_{\text{photon}} + \text{NEI}^2_0.
        \end{align}
        Combining Eqs.~\ref{eq:step_NEI} and~\ref{eq:meas_NEI} yields
        \begin{align}
            \label{eq:laser_cal}
            \text{NEI}^2_{\text{measured}} = \alpha I + \beta
        \end{align}
        where
        \begin{align}
            \label{eq:lasercal_fit_params0}
            \alpha =& 2h\nu_{\text{photon}}S(\nu) \\
            \label{eq:lasercal_fit_params1}
            \beta =& -\alpha I_0 - \text{NEI}^2_0.
        \end{align}
        The responsivity of the TES can be calculated from a fit to Eq.~\ref{eq:laser_cal} without knowledge of the total laser power incident on the detector. This calibration allows for a measurement of the absolute NEP from the measured NEI, even if the absolute NEI is not known, assuming that the NEP is proportional to the measured NEI with a constant multiplicative factor.

     \subsubsection{Measurements}  
\begin{figure}[h]
            \centering
            \includegraphics[width=3.1in]{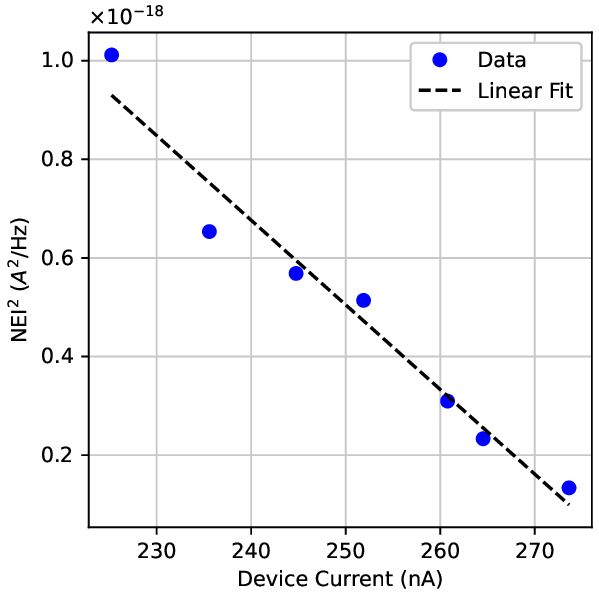}
            \caption{$\text{NEI}^2$ vs current with a linear fit. These data are extracted from the dataset presented in Fig.~\ref{fig:NEIvsI_raw}. The responsivity for this dataset is -70.1 nA/fW.}
            \label{fig:NEIvsI}
        \end{figure}

With the bias voltage fixed, a timestream of the TES current is measured for a range of laser powers. The TES current is the mean of the timestream data. The NEI is calculated from an FFT of the timestream data and is averaged over frequencies below the rolloff, but above any low-frequency excess, to capture the white noise value.
     An example of a TES timestream and the corresponding NEI is plotted in Fig.~\ref{fig:NEIvsI_raw}. The frequencies over which the NEI is averaged for this example are 0.3-3 Hz. The squared averaged NEI and the current corresponding to this example are plotted in Fig.~\ref{fig:NEIvsI}, with a linear fit. The fit parameter $\alpha$ is used to extract $S(\nu)$ using Eq.~\ref{eq:lasercal_fit_params0}. This calibration technique works for the TESs with and without absorbers. We see a factor of 5 higher absorption at 1550 nm in the TESs with absorbers.

\par To ensure that power fluctuations in the laser source (Thorlabs SLD) are not corrupting the measurement somehow, we repeated the experiment with a different source, a Thorlabs S3FC1550 distributed-feedback (DFB) laser\cite{thorlabs_dfb}.  We obtained the same results to within the measurement uncertainty.

    \subsection{Time Constant}
        \label{subsec:ac_time_constant}
        \begin{figure}[h]
            \centering
            \includegraphics[width=3.1in]{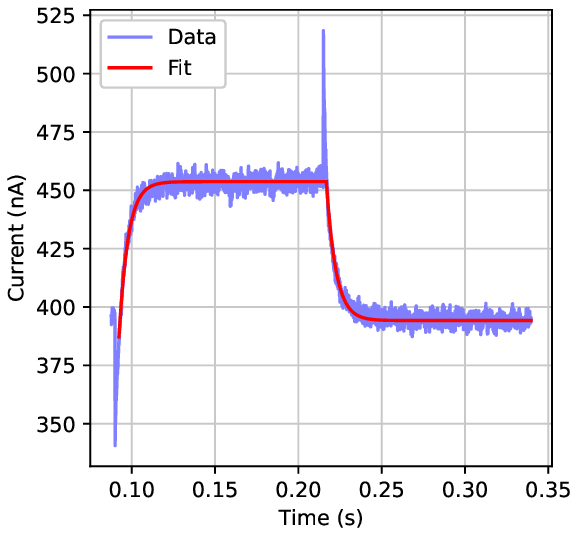}
            \caption{Example of time constant data and fits for ABS 1 (time constant = 6.4 ms). At each transition, the TES shows an short electrical transient  followed by the thermal response in the opposite direction, the latter being the relevant time constant.}
            \label{fig:tc}
        \end{figure}
        Time constants are measured from a detector response to a modulated electrical signal. A small square-wave modulation is applied to the TES bias voltage. The function generator time constant is sufficiently fast to measure the detector response. The response is fitted to an exponential function to extract the time constant. An example of the detector response to the modulated signal is plotted in Fig.~\ref{fig:tc} with exponential fits. 64 time constants are fitted and averaged for each TES (32 ascending and 32 descending). The averaged time constants for each of the four detectors are given in Table~\ref{tab:tc_table}. Our devices are slower than those from SRON \cite{Khosropanah_16,Ridder_16,Suzuki_16}.  Those measurements were obtained while biased deeper in the transition ($R\sim 0.25 R_n$), and were generally at bath temperatures of 40-60 mK, both of which increase the electrothermal feedback and, thus, the speed relative to the conditions of our measurements: biased at $R_n/2$ and a bath temperature of 100 mK. We may also have excess heat capacity on these devices. The fact that our devices with absorbers are slower than those without suggests additional heat capacity in the superconducting film, an aspect that is anomalous and may indicate excess heat capacity in the metallic components generally.
    \begin{table*}[t]
        \centering
        \begin{tabular}{c|c|c|c|c}
             TES      & Frequency (MHz) & $R / R_{\text{normal}}$ & Time constant (ms) & $S(\nu=0)$ (nA / fW)\\
             \hline
             ABS 1 &  2.346 970 & 0.538 & 6.4 & -70.1\\
             ABS 2 &  3.476 124 & 0.467 & 9.1 & -29.2\\
             NO ABS 1 &  2.423 641 & 0.449 & 3.7 & -51.1\\
             NO ABS 2 &  4.273 528 & 0.526 & 4.0 & -14.8\\
        \end{tabular}
        \caption{AC configuration resonance frequencies and averaged time constants for four measured TESs. ABS refers to a TES with an absorber, and NO ABS refers to a TES without an absorber. $R / R_{\text{normal}}$ is the fraction of the normal resistance at which the detector is biased for both noise and time constant measurements. The responsivities presented here are calibrated with the laser.}
        \label{tab:tc_table}
    \end{table*}

    \subsection{AC Setup NEP}
        \label{subsec:ac_nep}
\begin{figure}[h]
            \centering
            \includegraphics[width=3.1in]{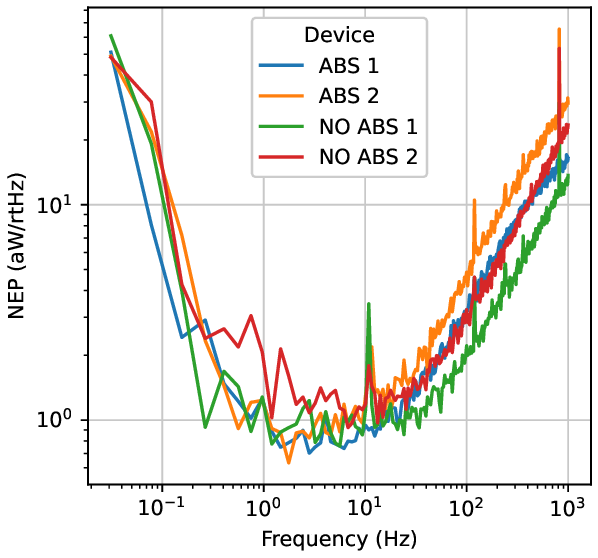}
            \caption{Calculated NEPs of four TESs, with the frequency response from Eq.~\ref{eq:S_freq_dependence} and the  measured time constants from Table~\ref{tab:tc_table}.}
            \label{fig:NEP}
        \end{figure}

        The responsivity is calculated using the photon-noise calibration method described in Sec.~\ref{subsec:photon_noise_derivation}. The responsivities are presented in Table~\ref{tab:tc_table}, along with the time constants that are used to calculate the frequency dependence of the responsivities. The spread in responsivities is an artifact of the laser calibration: the responsivities include the frequency-dependent gain of the system, an aspect that is removed from the NEP through the calibration. The calculated NEPs for the four TESs are shown in Fig.~\ref{fig:NEP}. The NEPs show a white noise of $1\times10^{-18}$ and rolloffs of 30 Hz set by the measured time constant.

\section{DC Setup Measurements}
    \label{sec:dc_measurements}
    As a check on the AC measurements and the inference from the near-IR shot noise, we employed the well-understood DC bias setup (Sec.~\ref{subsec:dc_setup}) to both directly measure the TES properties and to repeat the laser shot-noise calibration measurement.  For the DC measurements, we focused on the single device labeled ABS1 in Sec.~\ref{sec:ac_measurements}.

    \subsection{Current vs voltage }
        \label{subsec:dc_iv}
        \begin{figure}[h]
            \centering
            \includegraphics[width=3.1in]{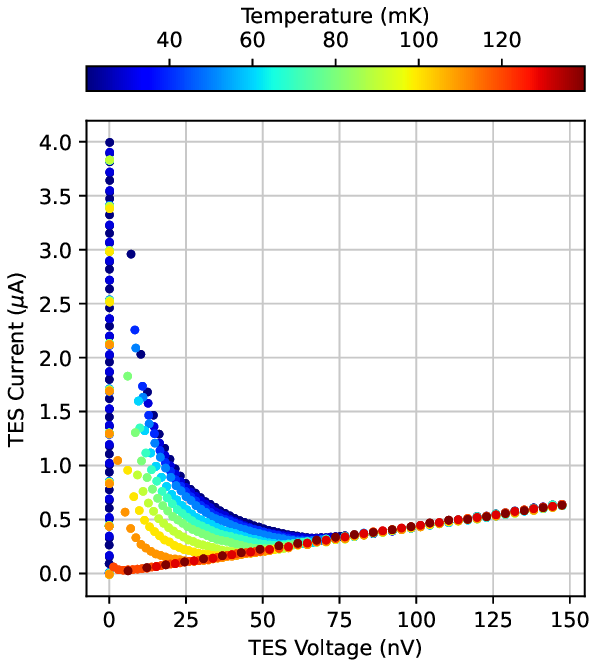}
            \caption{TES current vs voltage "load" curves in a DC mode.}
            \label{fig:IvV}
        \end{figure}
        \begin{figure}[h]
            \centering
            \includegraphics[width=3.1in]{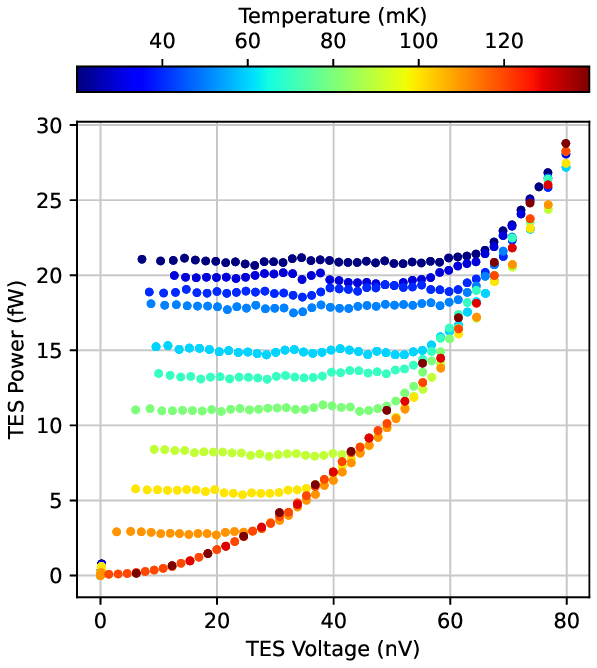}
            \caption{TES power vs voltage.}
            \label{fig:PvV}
        \end{figure}
        The total current through the parallel combination of the shunt resistor and TES is defined as $I_{\text{bias}}$. $I_{\text{bias}}$ is calculated from the applied function generator voltage using the circuit model in Fig.~\ref{fig:dc_circuit}. The TES current $I_{\text{TES}}$ is measured through the SQUID. For these DC measurements, the SQUID is operated with feedback, and the current calibration is confirmed to match the datasheet using the built in generator. The voltage is swept from high to low for multiple bath temperatures, and the TES current is plotted vs the TES Voltage in Fig.~\ref{fig:IvV}. The TES power is calculated from $P_{\text{TES}}=I_{\text{TES}}V_{\text{TES}}$ and plotted vs the TES voltage in Fig.~\ref{fig:PvV}. The resistance of the TES is calculated from $R_{\text{TES}}=R_{\text{shunt}}\left[I_{\text{bias}}/I_{\text{TES}}-1\right]$. The ratio $I_{\text{TES}}/I_{\text{bias}}$ is 0.977 for the superconducting branch, and 0.008 47 for the normal branch, corresponding to a negligible series resistance and a normal resistance of $211\;\rm m\Omega$.

    \subsection{Resistance vs temperature}
\begin{figure}[h]
            \centering
            \includegraphics[width=3.1in]{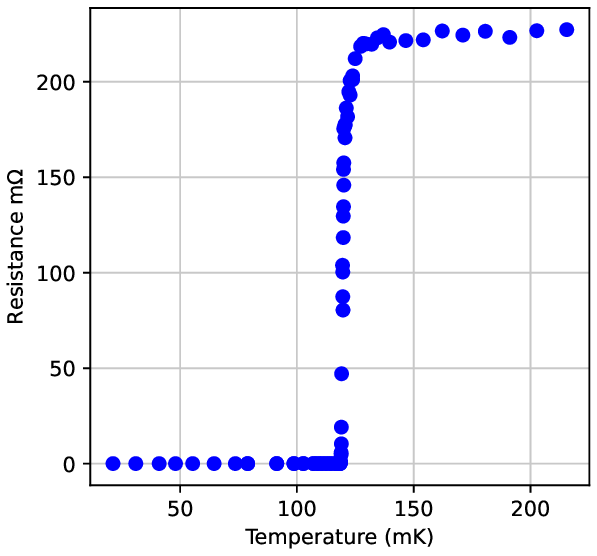}
            \caption{TES resistance vs temperature.}
            \label{fig:rvt}
        \end{figure}

        For a range of temperatures, the TES resistance can be measured from the TES voltage vs current. For each temperature, a linear fit to current vs voltage (in the regime where heating is negligible) determines the resistance at each temperature. The resistance vs temperature is presented in Fig.~\ref{fig:rvt}.

    \subsection{Thermal Conductance}
        \label{subsec:thermal_conductance}
        The temperature dependence of the thermal conductance $G=dP/dT$ is modeled as\cite{kenyon_progress_2006}
        \begin{align}
            \label{eq:G_T}
            G =& G_c \left(\frac{T_{\text{bath}}}{T_c}\right)^\beta.
        \end{align}
        \begin{figure}[h]
            \centering
            \includegraphics[width=3.1in]{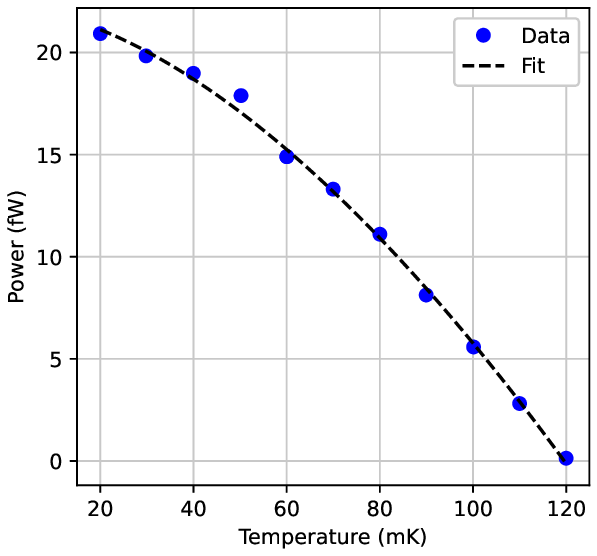}
            \caption{TES power vs temperature on transition with a fit to Eq.~\ref{eq:pvt}.}
            \label{fig:pvt}
        \end{figure}
        The power vs temperature relation can be calculated by integrating $dP=GdT$ with $G(T)$ defined in Eq.~\ref{eq:G_T}. The integrated equation is
        \begin{align}
            \label{eq:pvt}
            P =& \frac{G_c}{(\beta+1)T_c^\beta}\left(T_c^{\beta+1}-T_{\text{bath}}^{\beta+1}\right) - P_{\text{dark}}.
        \end{align}
        The power vs temperature data are extracted from the TES current vs voltage curves by averaging the power for each temperature in the transition. The power is plotted in Fig.~\ref{fig:PvV}. The negative slope in some of the power vs voltage curves is due to a small error in the current calibration. This error is negligible for our purpose of calculating the thermal conductance. The data are fit to Eq.~\ref{eq:pvt}. The data and fit are presented in Fig.~\ref{fig:pvt}. The fit parameters are presented in Table~\ref{tab:Gfit}. Note that $P_{\text{Dark}}$ is not a reliable measurement of the actual dark power because an accurate fit to $P_{\text{Dark}}$ requires a dark measurement of $T_c$ and the measurement of $T_c$ was under the same loading conditions as the power vs temperature data.
        \begin{table}[h]
            \centering
            \setlength{\tabcolsep}{0.2in}
            \begin{tabular}{c|c|c}
                \multicolumn{1}{c}{$\beta$} & \multicolumn{1}{c}{$G_c$ (fW/K)} &  \multicolumn{1}{c}{$P_{\text{Dark}}$ (fW)}\\
                \hline
                0.682 & 313 & 0.1411\\
                \hline
            \end{tabular}
            \caption{Power vs temperature fit parameters.}
            \label{tab:Gfit}
        \end{table}

\subsection{Photon-Noise Calibration}
\begin{figure}[h]
            \centering
            \includegraphics[width=3.1in]{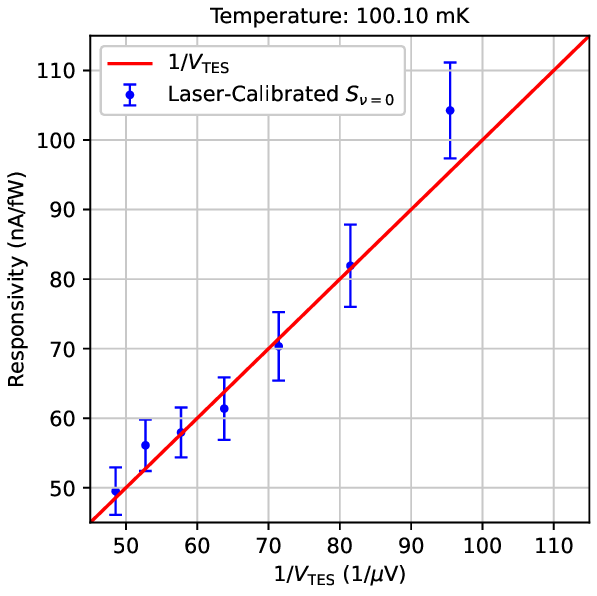}
            \caption{DC responsivity: comparison of electrical responsivity estimated from the applied voltage and the circuit, and via the photon-noise-calibration technique. $V_{\text{TES}}$ is calculated from the circuit model in Fig.~\ref{fig:dc_circuit}, where $R_{\text{shunt}}$ is set to the nominal shunt resistor value of $1.8\;\text{m}\Omega$. The vertical error bars are propagated from the fits to $\text{NEI}^2$ vs $I$.}
            \label{fig:NEIvsI_DC}
    \end{figure}

    The photon-noise calibration was repeated with the DC setup. The DC electrical responsivity $S(\nu=0)$ is equal to the inverse of the TES voltage, which is calculated from the circuit model in Fig.~\ref{fig:dc_circuit}. The photon-noise calibration was repeated at several bias voltages at 100.10 mK. The results are plotted in Fig.~\ref{fig:NEIvsI_DC} with the electrical responsivity. The error bars are propagated from the uncertainty in the linear fits to $\text{NEI}^2$ vs $I$.

\subsection{NEP Common-Mode Reduction}
    In order to determine whether the 1/f noise is intrinsic to the detector or due to the measurement setup, an array was tested with a time-domain readout system at the Goddard Spaceflight Center. The time-domain system allows for common-mode reduction in the noise spectra of multiple detectors that were measured simultaneously. Four detectors without absorbers were measured.
    \subsubsection{Experimental Setup}
        The laboratory Dewar is a custom-built cryostat designed at Goddard. It is a combination of parts from Simon Chase Cryogenics (4He sorption Cooler), High Precision Devices (ADR) and CryoMech ( PT-407). A custom-built set of readout electronics is used to read and run all of the thermometers and control the PID loop for the ADR at a level of less than $200\;\text{nK}/\sqrt{\text{Hz}}$.
        \par Clocking and grounding are controlled and synchronized for electrical noise performance. Vibrational dampers are used to control acoustic noise that can be introduced from floor and air vibration.
        \par The thermal and electrical control provides a clean environment for bolometer testing. To provide a regime that makes it possible to test low-noise bolometers, a specially designed package was developed at Goddard. This package utilizes superconducting thermal filters that isolate the SQUID readout chips (the time-domain SQUID multiplexers and series arrays were provided by NIST/Boulder) from the detector chip under test. Both regions in the package are darkened with a special EMI/Thermally black paint developed at Goddard\cite{chuss_cryogenic_2017} and then closed out with a metal to metal stepped edge to block external light from getting in. The multiplexed detector signals were read out using the Multi Channel Electronics (MCE) from UBC\cite{battistelli_functional_2008}.

    \subsubsection{Measurements and Data Reduction}
        \par The data for the plots were taken with a base temperature of 88 mK. The data were taken at night to minimize interference from other sources. There were three runs with different bias settings, each operating the detector on a different place on the transition. Data were taken for $\sim 800\;\text{s}$ each of those runs. The sample rate is 50 MHz and there were 4 pixels (rows) with a dead time of 4.8 $\mu$s (240 samples)  for settling and a dwell time of 1.2 $\mu$s (60 samples) on each row. Thus, there were 33.33 Msamples for each row/run.
        \par In two of these row/runs a single jump was noted.  The data before the jump were averaged and the mean was subtracted.  The data after the jump were also averaged and the mean was subtracted. Then, for each row/run, a straight line fit (offset and slope) was removed.  The slope is minimal and probably not important.  There is an arbitrary offset set in at the time the feedback is locked. The data were then Fourier transformed, the absolute value was applied, and half the results were binned into frequency bins. The remaining half is a mirror image and, thus, redundant. This forms a standard power spectrum. The gain was determined by the voltage setting on the TES and the counts to generate a $\Phi_0$ in the feedback. The current to power responsivity is given as the standard $V_{\text{bias}}$. The shunt resistor is $250\;\mu\Omega$.
        \par
\begin{figure}[h]
            \centering
            \includegraphics[width=3.1in]{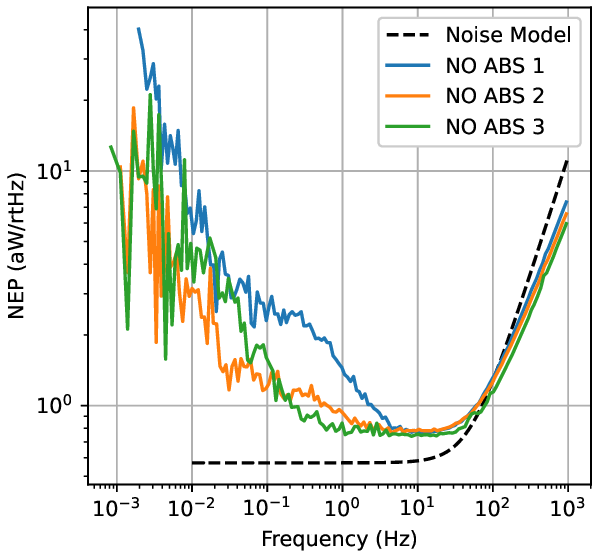}
            \caption{Calculated NEP of three TESs without absorbers, with the common-mode noise removed. The noise model includes the G noise, the SQUID noise, the Johnson noise, and the response rolloff.}
            \label{fig:goddard_nep}
    \end{figure}

 The resulting power spectra were then averaged (RMS) into equal log bins in frequency.  Those at the lower end of the spectrum are still individual points in the FFT. The frequencies were averaged to get an average frequency for each bin. This has the effect of smoothing the data at high frequency although some sharp lines still appear. The lowest frequency is limited by the duration ($800\;\text{s}$) of the data. The PID controller was turned on but was not properly tuned for the
        temperature setting.  We show results for three of the devices (there was a data glitch during the measurements of the fourth device). The noise data are presented in Fig.~\ref{fig:goddard_nep}. The white noise level is $0.8\;\text{aW}/\text{Hz}^{1/2}$, and the single-pole rolloff is at 50 Hz (3.2 ms time constant). NO ABS 2 and 3 have a 1/f knee below 0.1 Hz, and NO ABS 1 has a higher 1/f knee of below 1 Hz. The 1/f noise is reduced from the measurements in Sec.~\ref{subsec:ac_nep} by up to a factor of 10.
        \par The noise model in Fig.~\ref{fig:goddard_nep} includes the G noise, the Johnson noise, and the SQUID noise. The G noise is $\text{NEP}_{\text{G}}=\sqrt{4kT^2G}=0.53\;\text{aW}/\sqrt{\text{Hz}}$, where G is measured in Sec.~\ref{subsec:thermal_conductance}. The SQUID noise was measured using an off-resonance bias, and matches the Magnicon datasheet. The SQUID noise divided by the average responsivity of the TESs is $0.08\;\text{aW}/\sqrt{\text{Hz}}$. The Johnson noise is calculated from $\text{NEI}_{\text{Johnson}} = \sqrt{\frac{4kT}{R}}$, where $T$ and $R$ are the operating temperature and resistance of the TESs. The average value of the Johnson noise divided by the average responsivity is $0.19\;\text{aW}/\sqrt{\text{Hz}}$. The sum of the G, SQUID, and Johnson noise is $0.57\;\text{aW}/\sqrt{\text{Hz}}$.

\section{Summary}
    \label{sec:summary}
    We present development and testing of low-thermal-conductance transition-edge-sensed (TES) bolometers for far-IR space astrophysics.  Our new wet-release process is compatible with large arrays of devices with 1-mm long legs on 1/4-micron silicon nitride, and yields an improved speed of response (3--10~ms) relative to previous low-G devices developed at the JPL Microdevices Lab (10-30 ms)\cite{Beyer12,Beyer14,Kenyon_14}.  However, in a careful calibration in both AC and DC bias schemes and multiple experimental setups, we find that the thermal conductances (Gs) are ten times higher than previously measured with the same leg geometries in the old dry-release process.  We do note that a similar wet-release approach used for higher-G, higher-temperature devices (not presented here) confirms an increase in speed with fixed G relative to the dry process; therefore, the heat capacity improvements are real.  The higher G in our low-G wet-release devices may be due to a reduction in phonon scattering along the legs due to the smoothness associated with the wet release.
    \par In characterizing these devices, we have developed a near-IR photon-shot-noise calibration technique which offers a rapid and unambiguous measure to total power on the TESs.  While not a replacement for far-IR optical characterization, the near-IR technique is very convenient for assessing electrical properties.  This is especially useful when the bias transfer is uncertain, as is often the case in AC bias systems.  It works in the regime in which the power to current responsivity is constant, the bulk of the operating regime for TES devices. In future experiments, stray light could be reduced using a simple near-IR bandpass filter.
    \par The time-domain multiplexing system at Goddard yields better sensitivities than the AC setup, a fact we attribute to improved filtration and stray light blockage, rather than a fundamental difference in the approach.  Through a careful 1/f reduction, the noise is further reduced to an electrical noise-equivalent power as low as $0.8\;\text{aW}/\text{Hz}^{1/2}$.  This is only modestly  larger than the $0.57\;\text{aW}/\text{Hz}^{1/2}$, which is expected from the fluctuations in the thermal transport (the G noise), the Johnson noise, and the SQUID noise. The floor is very white, suggesting that it may be due to stray light.  We measure a 1/f knee frequency as low as 0.1 Hz, ample for large-scale mapping experiments envisioned for the cryogenic space missions.

\begin{acknowledgments}
The research was carried out at the Jet Propulsion Laboratory, California Institute of Technology, under a contract with the National Aeronautics and Space Administration (80NM0018D0004). This work was funded by the NASA (Award No. 141108.04.02.01.47)—to Dr. C. M. Bradford. We would like to thank Jonathan Hunacek and Brian Steinbach for helpful discussions.
\end{acknowledgments}
\section*{Data Availability Statement}
The data that support the findings of this study are available from the corresponding author upon reasonable request.

\bibliography{bibliography.bib}

\end{document}